\documentclass[12pt,preprint2]{aastex}

\usepackage[margin=1.0 in]{geometry}
\usepackage{times}
\usepackage{amsmath}
\usepackage{rotating,tabularx}

\newcommand{\bc}{\begin{center}}
\newcommand{\ec}{\end{center}}
\newcommand{\be}{\begin{equation}}
\newcommand{\ee}{\end{equation}}
\newcommand{\bfig}{\begin{figure}}
\newcommand{\efig}{\end{figure}}

\newcommand{\Ms}{M$_\odot$}
\newcommand{\mm}{$\mu$m}

\extrafloats{100}

\received{?}
\revised{?}
\accepted{?}
	
\begin{document}

\title{High-resolution Spectra and Biosignatures of Earth-like Planets Transiting White Dwarfs}
\author{Thea Kozakis, Zifan Lin, Lisa Kaltenegger}
\affil{Carl Sagan Institute, Cornell University, Ithaca, New York, USA}
\date{}


\begin{abstract}
With the first observations of debris disks as well as proposed planets around white dwarfs, the question of how rocky planets around such stellar remnants can be characterized and probed for signs of life becomes tangible. White dwarfs are similar in size to Earth and have relatively stable environments for billions of years after initial cooling, making them intriguing targets for exoplanet searches and terrestrial planet atmospheric characterization. Their small size and the resulting large planet transit signal allows observations with next generation telescopes to probe the atmosphere of such rocky planets, if they exist. We model high-resolution transmission spectra for planets orbiting white dwarfs from as they cool from 6,000-4,000~K, for i) planets receiving equivalent irradiation to modern Earth, and ii) planets orbiting at the distance around a cooling white dwarf which allows for the longest continuous time in the habitable zone. All high-resolution transmission spectra will be publicly available online upon publication of this study and can be used as a tool to prepare and interpret upcoming observations with JWST, the Extremely Large Telescopes as well as mission concepts like Origins, HabEx, and LUVOIR.
\vspace{0cm}

\end{abstract}

\vspace{3cm}

\maketitle

\section{Introduction}
The first discovery of a planetestimal orbiting a white dwarf (WD) in 2015 \citep{vand15} has been joined by other recent discoveries \citep{mans19,gans19}, which postulate the first indirect detection of a planet orbiting a WD. The high occurrence rate of heavy metal WD pollution observed by many groups (e.g.\ \citealt{koes06,klei11,koes14,hame16,mala16}) indicate debris disks or planets around a high percentage of WDs of up to 50\% \citep{schr19}. K2 statistics constrain the rate of WD habitable zone (HZ) planet occurrence to $<$28\% \citep{vans18}. Several studies have used WD pollution to infer the dynamical evolution \citep{vera19,vera19b} and the composition of accreting planets (e.g.\ \citealt{swan19}), with \cite{baue19} finding WD pollution is sometimes consistent with several Earth masses of rocky debris accretion during the early stages of WD cooling, potentially implying close-in terrestrial planets.  WD stellar remnants are only slightly larger than Earth, with long cooling timescales, which give them a long-lived, stable WD HZ, which can provide temperature conditions for rocky planets of $\sim$8 billion years (Gyr) \citep{koza18}. Therefore rocky planets in the WD HZ are interesting planets both to search for as well as to characterize second-generation terrestrial planets.

Multiple studies have addressed the possibility of WD planet detection via transits (e.g.\ \citealt{agol11,loeb13,cort19}) and several searches are already underway (e.g.\ \citealt{fult14,vera15,xu15,wall18, bell19,dame19}). The potential for habitable planets orbiting in the WD HZ during its cooling process (e.g.\ \citealt{agol11,barn13,koza18}) and the UV surface environment and its potential impact on surface life have been discussed by several teams (e.g.\ \citealt{mccr71,foss12,koza18}).
A first estimate for the strength of transmission spectral features for Earth-like planets around WDs has been explored by \cite{loeb13}, who used the modern Earth transmission spectrum in lieu of atmospheric models for Earth-like planets orbiting WDs. While this was a useful first approach, the different irradiation environment around a WD compared to around the Sun, especially in the UV, changes the atmospheric composition as well as the spectral features compared to modern Earth. We show the modern Earth transmission spectrum \citep{kalt09} in our transmission spectra figures for comparison. 

This paper provides the first high-resolution transmission spectra database for Earth-like planets orbiting WD from 6,000 to 4,000 K both for planets i) receiving Earth-analog irradiation from the WD at one point in its evolution, as well as ii) for a planet at a specific orbital distance, which would allow for the longest continuous time in the WD HZ during the WD's evolution (the atmosphere models are discussed in detail in \citealt{koza18}). During the WD cooling process planets in the WD HZ will experience a constantly decreasing overall incident flux, along with decreasing incident UV flux, which impact the planet's climate and atmospheric photochemistry (ibid). 
Section 2 describes our models, Section 3 presents our results, and Section 4 summarizes and discusses our findings.
\vspace{0.4cm}

\section{Methods}

\subsection{WD cooling and spectral models}

Newly formed WDs are extremely hot (up to 100,000 K), however they gradually cool over time due to a lack of an internal heat source. An average WD has cooled to 6,000 K after $\sim$2 Gyr, but then takes an additional $\sim$8 Gyr to reach 4,000 K \citep{berg01}, providing planets nearly twice Earth's lifetime in the continuous WD HZ (as discussed in detail in \citealt{koza18}). To explore WD planet evolution throughout their host's cooling, we model the photochemistry and climates of such planets using WD spectral models described in \cite{saum14} for WD hosts at 6,000, 5,000, and 4,000 K. The models assume pure hydrogen atmospheres for the average WD mass of 0.6 \Ms\ \citep{kepl16}. These WD spectra only show hydrogen lines above 5,000 K, and are essentially black bodies under 5,000 K, at which point hydrogen becomes neutral (ibid).

\vspace{0.4cm}

\subsection{Planetary atmospheric models and spectra}

To model planetary atmospheres and resulting spectra we use \emph{Exo-Prime} (see e.g.\ \citealt{kalt10b}) which couples a 1D climate code (based on \citealt{kast86,pavl00,haqq08}), a 1D photochemistry code (based on \citealt{pavl02,segu05,segu07}), and a radiative transfer code (based on \citealt{trau76,kalt09}). This code was designed for rocky planets and models temperature, chemical profiles, UV surface fluxes, and emergent and transmission spectra. Figure~\ref{photochem} and Table~\ref{model_summary} summarize the model parameters, temperature and chemical mixing ratios of the WD planet models described in detail in \cite{koza18}.

Using WD irradiation spectra (described above) as incoming irradiation, we divide the atmosphere into 100 parallel layers up to a pressure of 1 mbar. To model planets at the Earth-equivalent distance we scaled the integrated flux of the WD input stellar spectrum to the solar constant. After the models are run, we factor in limitations in the atmospheric depths we can probe due to refraction, which changes based on the geometry of the system.  Outgoing light rays must be parallel to reach a distant observer, thus rays that are bent strongly in dense, deep regions of a planetary atmosphere will not contribute to the observed signal. Due to this effect, an Earth-analog planet's atmosphere at 1AU orbiting a Sun-sized star can only be probed down to about 12.7 km above the planetary surface (e.g. \citep{betr14, macd19}), while for a planet around a WD, the atmosphere of a planet receiving Earth-analog irradiation can be observed down to 6.5 km above the surface \citep{macd19}. Therefore we cut our spectra off at the effective height equivalent to the system's lowest observable altitude.

All spectra are calculated at high resolution 0.01 cm$^{-1}$ covering wavelengths from 0.4 to 20 \mm\ and will publicly available online upon publication of this study. All transmission spectra are plotted at a resolution of $\lambda/\Delta\lambda$~=~700 in the figures for clarity.

\section{Results: Transmission spectra of Earth-like planets orbiting WDs}

\subsection{Transmission spectra of planets orbiting WDs at the Earth-equivalent distance}

\begin{figure*}[h!]
\centering
\includegraphics[scale=0.8]{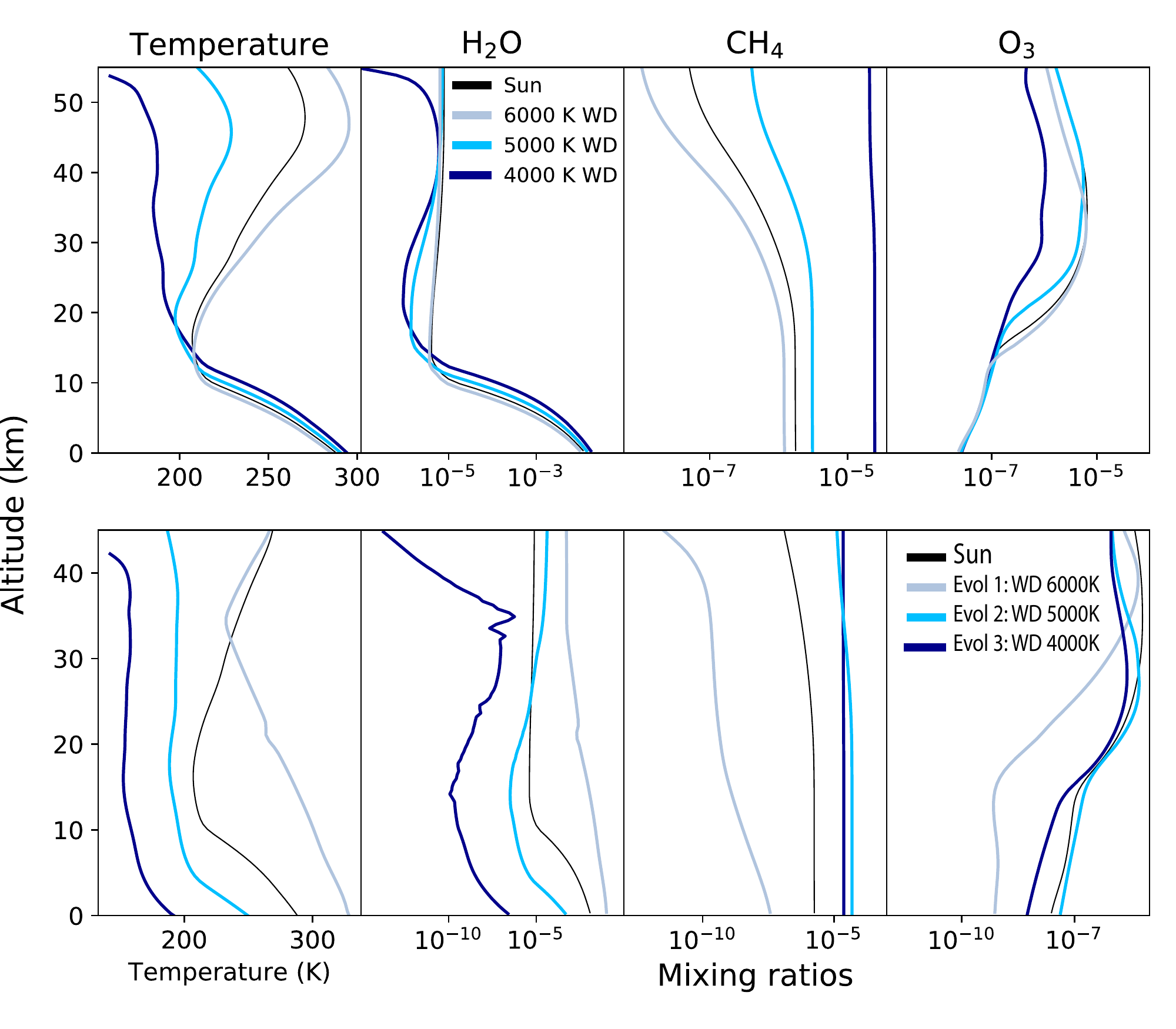}
\caption{Temperature and mixing ratio profiles of Earth-like planet models orbiting WDs (top) at the Earth-equivalent distance and (bottom) on a specific orbit which allows for the maximum time in the WD HZ as the WD cools, shown for 3 evolutionary stages. \label{photochem}}
\end{figure*}

\begin{table*}[t!]
\begin{center}
\caption{Model summary for Earth-like planets orbiting WD hosts receiving i) Earth-equivalent irradiance and ii) on a specific orbit which allows for the maximum time in the WD HZ as the WD cools, shown for 3 evolutionary stages \citep{koza18}.  We also compare the semimajor axes, $a$, of these planets to the Roche limit for a rigid satellite, $r_{Roche}$, which for a WD with a mass of of 0.6 M$_\odot$ and a radius of 0.0128 R$_\odot$ corresponds to a value of 0.67 R$_\odot$.   \label{model_summary}}
\begin{tabular}{lcccc}
Stellar host T$_{\footnotesize \mbox{eff}}$ \hspace{1cm}& S$_{\footnotesize \mbox{eff}}$ \hspace{0.5cm}& $a/r_{Roche}$  &Planet surface T$_{\footnotesize \mbox{eff}}$& Ozone column depth \\
(K) & &  & (K) &  (cm$^{-2}$) \\
\hline
Present day Earth \hspace{0.5cm}& 1.00 & - & 288.2 & 5.4$\times10^{18}$\\
\hline
6000 & 1.00 & 		4.5		& 285.6 &		5.7$\times10^{18}$\\
5000	 & 1.00 &		3.1		& 290.8 &		3.9$\times10^{18}$\\
4000 & 1.00 &		2.0		& 294.3 &		1.8$\times10^{18}$\\
\hline
6000	&	1.34	&	3.9		& 328.4	&	1.7$\times 10^{18}$ 	\\
5000	&	0.64	&	3.9		& 249.6	&	4.4$\times 10^{18}$ 	  \\
4000	&	0.26 &	3.9		& 191.9	&	9.6$\times 10^{17}$   \\
\hline
\label{planets}
\end{tabular}
\end{center}
\end{table*}

As a WD cools, its UV flux steadily decreases, causing significant changes in our model planets' atmospheric photochemistry, which is highly sensitive to the amount of incoming UV radiation (see Figure~\ref{photochem}) (as discussed in detail in \citealt{koza18} and shortly summarized here).  Ozone (O$_3$) production requires high energy UV photons with $\lambda <$ 240 nm, causing production rates to decrease for planets orbiting cooler WDs with less incident UV, and lowering the atmosphere's ability to shield the surface from UV radiation.  Note that shortward of about 200 nm, absorption by atmospheric CO$_2$ filters out biologically harmful UV flux (see e.g. discussion in \citealt{koza18,omal19}).

The decrease of ozone leads to a decrease of its byproduct hydroxyl (OH), which is one of the main sinks for methane (CH$_4$). Methane additionally undergoes significant depletion during photolysis in high UV environments. Cooler WDs also emit a larger percentage of their light at longer wavelengths, resulting in more efficient planetary surface heating and higher planetary surface temperatures for cooler hosts. Thus with the evolution and cooling of the WD host, atmospheres of Earth-like planets show less ozone, more methane, higher surface temperatures and decreasing temperature inversion in the planetary model atmospheres for similar overall incident flux from cooler WD hosts. Details of the model parameters are shown in Table~\ref{model_summary} and temperature and chemical profiles are shown in Figure~\ref{photochem} for both i) planets orbiting at the Earth-equivalent distance, and ii) a planet on the orbit that allows the maximum time in the HZ during WD cooling.

\begin{figure*}[h!]
\centering
\includegraphics[scale=0.5]{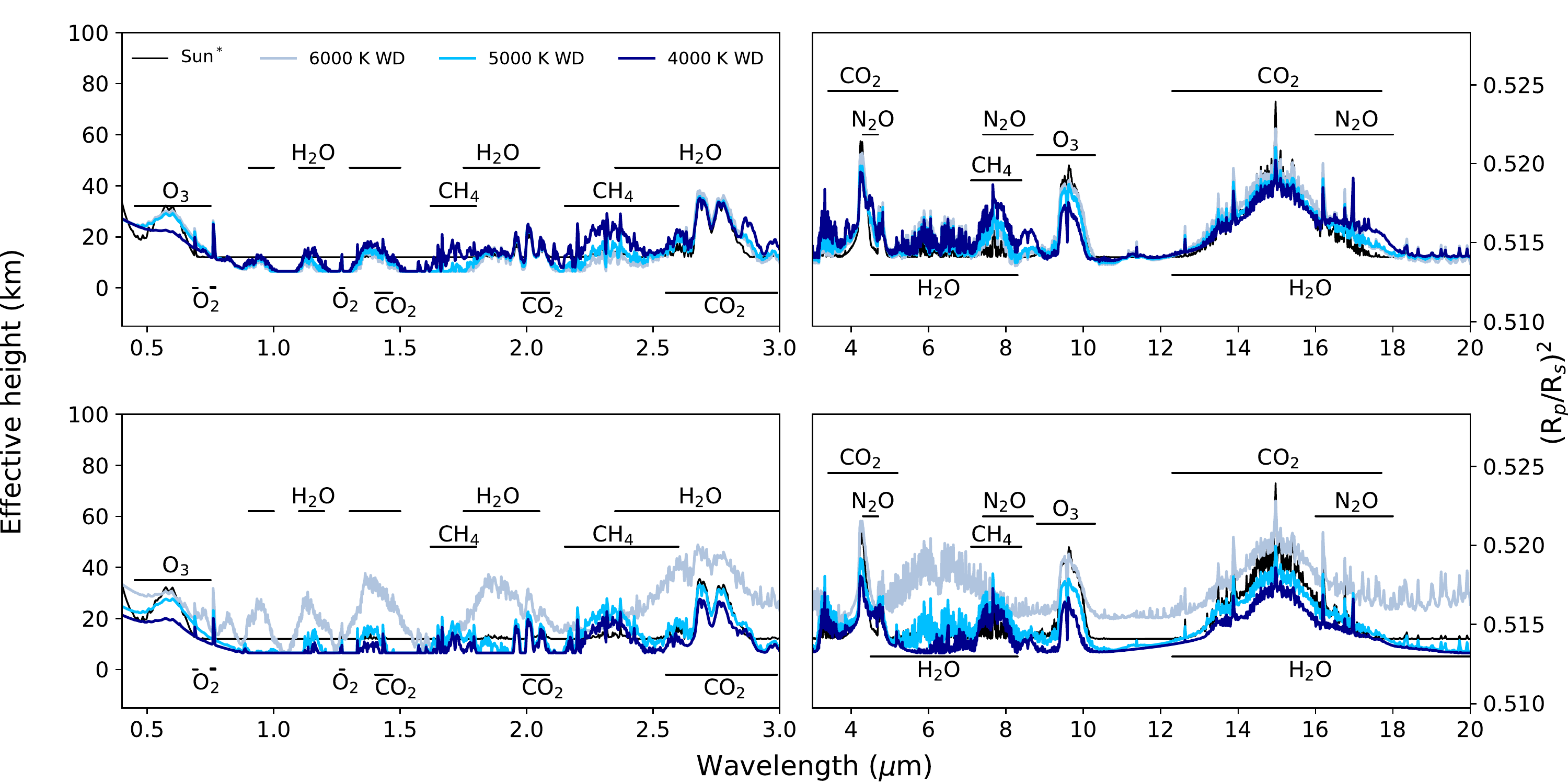}
\caption{Transmission spectra for (top) planets orbiting at the Earth-equivalent distance and (bottom) a planet with a static semimajor axis 3.9 times the Roche limit throughout WD cooling for 3 evolutionary stages (4,000, 5,000, and 6,000 K WD) both in terms of effective height (left axis) and (R$_p$/R$_s$)$^2$ (right axis). Major spectral features are labeled. The Earth-Sun system is shown for comparison. Note that the transmission spectra of Earth is multiplied by 1.65$\times$10$^4$ to show on the same contrast ratio scale for comparison. \label{all_spectra}}
\end{figure*}

\begin{figure*}[h!]
\centering
\includegraphics[scale=0.7]{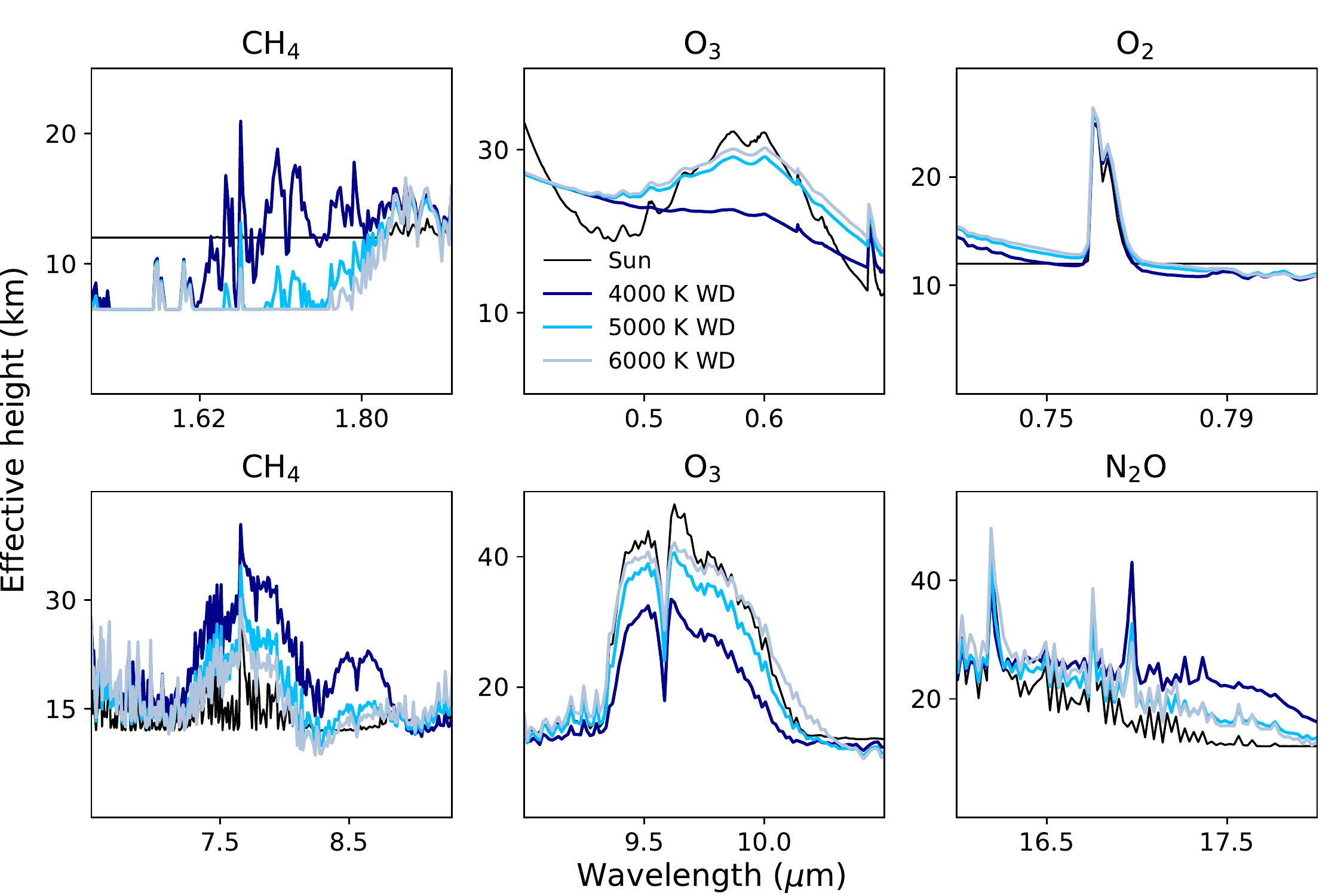}
\caption{The most prominent spectral features for potential biosignatures of planets orbiting at the Earth-equivalent distance in the (top) visible to near-IR for O$_3$ at 0.6 \mm, CH$_4$ at 1.7 \mm\ and O$_2$ at 0.76 \mm\, and in the IR (bottom) for O$_3$ at 9.6\ mm, CH$_4$ at 7.6 \mm\ and N$_2$O at 17 \mm. The black line shows the modern Earth-Sun system for comparison.\label{multi}}
\end{figure*}

Figure~\ref{all_spectra} shows our transmission spectra calculated using these atmospheric models from \cite{koza18}, with zoomed in biologically relevant features shown in Figure~\ref{multi}.  A unique differences between transmission spectra of WDs versus main sequence stars is the large difference in (R$_p$/R$_s$)$^2$,where R$_p$ is the planet's effective radius, and R$_s$ is the radius of the stellar host. This quantity determines the depth of the transit, with larger (R$_p$/R$_s$)$^2$ values creating a larger signal.  If we were to consider an Earth-sized planet around a Sun-sized star (R$_s$ = 1 R$_\odot$) versus a WD (R$_s$ = 0.00128 R$_\odot$), the corresponding (R$_p$/R$_s$)$^2$ values for the planet without considering any atmospheric absorption would be 8.4$\times10^{-5}$ and 5.1$\times10^{-1}$, respectively. With similar effective atmospheric heights for both cases, the contrast ratio is 4 orders of magnitudes larger for such planets orbiting WDs. In Figure~\ref{all_spectra} the Earth-Sun transmission spectra were multiplied by 1.65$\times10^4$ to show on the same contrast ratio scale as the WD planet model spectra.  The transmission spectra in Figure~\ref{all_spectra} are shown at a resolution of $\lambda/\Delta\lambda$ =  700 for clarity.

H$_2$O: Although the higher surface temperatures for the cooler WD hosts result in increased water abundance in the upper atmosphere than for hotter WD hosts, the spectral features are dominated by the lower atmosphere, thus water features are stronger for planet models orbiting hotter WDs, because larger amounts of ozone protect the water from depletion via photolysis (see Figure~\ref{photochem}). The strongest H$_2$O transmission spectra absorption features can be seen in Figure~\ref{all_spectra} at 0.9, 1.4, 1.9, 5 and 20 \mm. 

O$_2$: The strongest O$_2$ features in the transmission spectra models in the modeled wavelength range from 0.4 to 20 \mm\ is at 0.76 \mm\ (see Figure~\ref{multi} for a detailed view). All model atmospheres assume a O$_2$ mixing ratios of 0.21, leading to similar strength of the oxygen absorption line.  The differences are due to overlapping features.

O$_3$: The two strongest O$_3$ spectral features in our transmission spectra models are at 0.6 and 9.6 \mm (seen in Figure~\ref{multi}). The spectral features increase for hotter WD hosts due to higher production rates of ozone. 

CH$_4$: The strongest CH$_4$ absorption features in transmission are at 1.7 and 7.6 \mm\ (seen in Figure~\ref{multi}). CH$_4$ absorption features are stronger in the atmosphere for the coolest WD case, where its concentration increases because of the lower UV environment causing less depletion via photolysis as well as less reactions with OH, a byproduct of ozone, which is also reduced in the atmosphere for cool WDs. 

CO$_2$: The strongest CO$_2$ absorption features in transmission are at 2.0, 2.7, and 15 \mm. CO$_2$ is set to a fixed mixing ratio of modern Earth concentration for all models and the absorption features shows comparable strength for all model planets.

N$_2$O: The strongest N$_2$O absorption features in transmission are at 4.4, 8, and 17 \mm\ (see Figure~\ref{multi} for details on the 17 \mm\ feature). N$_2$O is depleted similiarly to CH$_4$ around hotter WD hosts due to increased photolysis and reactions via OH.

The most prominent spectral features of potential biosignatures in the visible to NIR in transmission are O$_3$ at 0.6 \mm, CH$_4$ at 1.7 \mm\ and O$_2$ at 0.76 \mm, and in the IR O$_3$ at 9.6 \mm, CH$_4$ at 7.6 \mm, and N$_2$O at 17 \mm. These spectral features are shown in detail in Figure~\ref{multi} (top: visible to NIR 0.4 to 3 \mm, bottom IR 3 to 20 \mm) for planets orbiting at the Earth-equivalent distance.

\subsection{Transmission spectra of rocky planets in the WD HZ throughout white dwarf cooling}

An average WD takes about $\sim$8 Gyr to cool from 6,000 K to 4,000 K \citep{berg01}, providing a stable, continuous WD HZ during that time (discussed in detail in \citealt{koza18}). A planet orbiting at a semimajor axis 3.9 times the Roche limit from its WD host would spend $\sim$6 Gyr in the conservative WD HZ, and $\sim$8.5 Gyr in the empirical WD HZ (ibid) using empirical HZ limits based on early Mars and recent Venus irradiation \citep{kast93}. Such a planet would initially (Evolution stage 1) receive 134\% of modern Earth's flux from its 6,000 K WD host, which reduces to 64\% for a WD which cooled to 5,000 K (Evolution stage 2), and 26\% of modern Earth's flux when the WD cools to 4,000 K (Evolution stage 3). This change in irradiation is similar to a planet at an orbital distance between Venus and modern Earth, to a Mars orbit in our solar system. 

\begin{figure*}[h!]
\centering
\includegraphics[scale=0.7]{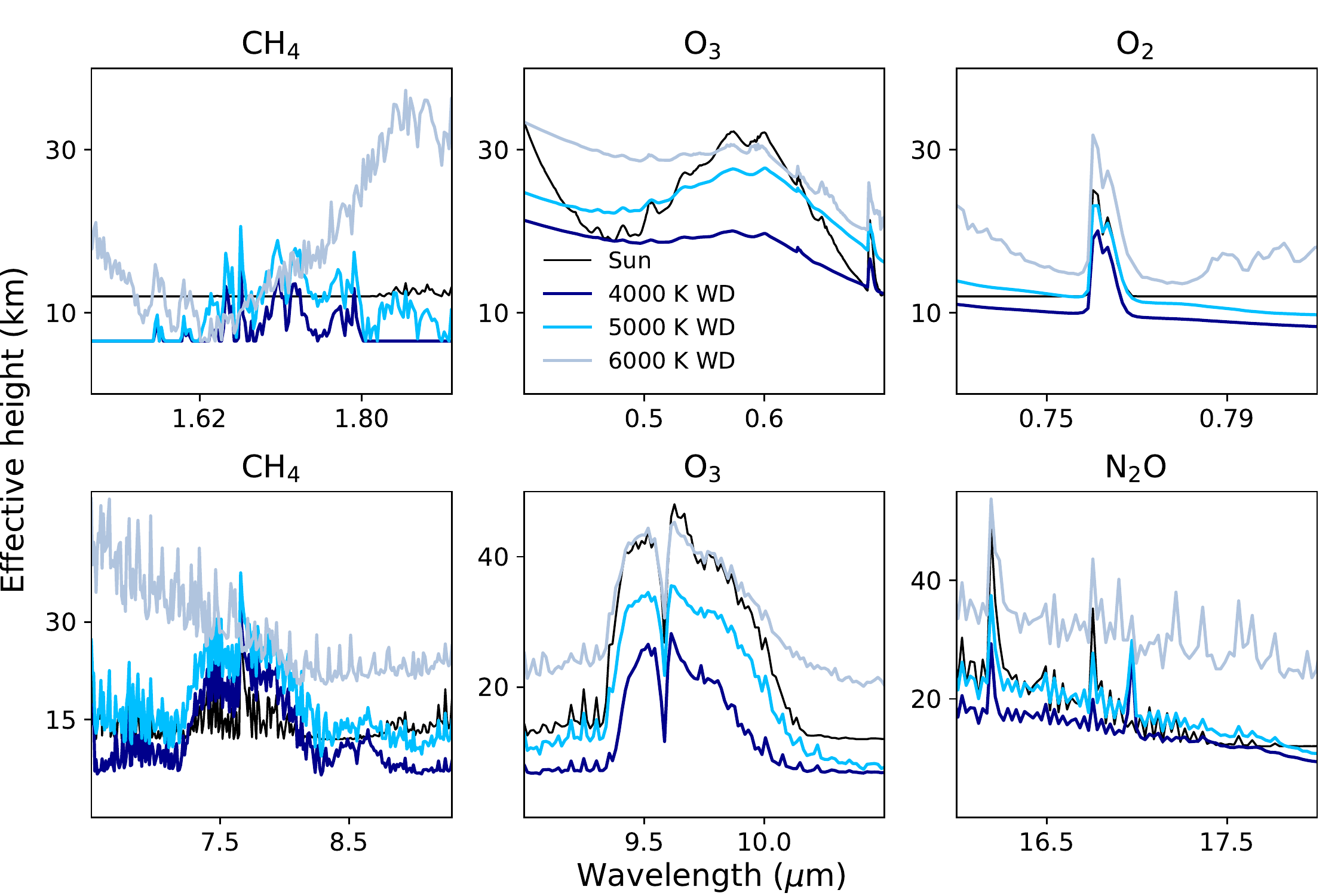}
\caption{The most prominent spectral features for potential biosignatures shown in the (top) visible to near-infrared for O$_3$ at 0.6 \mm, CH$_4$ at 1.7 \mm\ and O$_2$ at 0.76 \mm, and in the IR (bottom) for O$_3$ at 9.6\ mm, CH$_4$ at 7.6 \mm\ and N$_2$O at 17 \mm\ for a rocky planet at an orbital distance, which allows for maximum time in the WD HZ shown at 3 points throughout the WDs cooling process (at 6,000, 5,000, and 4,000 K WD host effective surface temperature). The modern Earth-Sun system spectra is shown for comparison.\label{multi_evol}}
\end{figure*}

During this WD cooling process the amount of incident UV flux upon the planet steadily decreases, changing the planet's atmospheric chemistry as well as its UV surface environment. Details of the model parameters from \cite{koza18} are shown in Table~\ref{model_summary} and temperature and mixing ratio profiles are shown in the bottom row of Figure~\ref{photochem}. We summarize the results here to link them to the spectral features shown in Figure~\ref{all_spectra}, with specific biologically relevant features shown in Figure~\ref{multi_evol}.

At the first stage in the modeled evolution of such a WD planet, for a 6,000 K WD host such a planet receives high UV flux, causing large ozone production as well as high rates of photolysis in the model atmospheres. O$_3$, CH$_4$, and N$_2$O are significantly depleted via photolysis. The high total incident flux causes high surface temperatures and thus high amount of water vapor throughout the atmosphere.

At the second stage of the modeled evolution, for a 5,000 K WD, with a decreasing incident UV and overall flux the planet shows a substantial ozone layer because of reduced O$_3$ depletion via photolysis. The concentration of CH$_4$ and N$_2$O also increase compared to the first stage because of reduced depletion via photolysis.

At the third stage, for a 4,000 K WD host, the reduced incident flux causes cold surface temperatures and very low photolysis rates. Only a small amount of ozone is produced in our models and there is very little depletion via photolysis for all chemical species. Due to the low surface temperature, water concentration is also lower. Note that we did not consider in our models that a similar geological cycle to Earths carbonate-silicate cycle could increase the CO$_2$ concentration in the atmosphere of such colder planets, maintaining warm surface temperatures than shown in tour models (see \citealt{koza18}). \

Most notable in Figure~\ref{all_spectra} are the differences in H$_2$O absorption, caused by the decrease of H$_2$O for a cooling WD host. High ozone production rates in the first evolution stage for the 6000~K WD produce stronger ozone features in evolution stage 1. Methane shows stronger absorption features for the later evolution stages 2 and 3, the two cooler WD cases, where photolysis rates are low.

The most prominent spectral features of potential biosignatures in the visible to NIR in transmission are O$_3$ at 0.6 \mm, CH$_4$ at 1.7 \mm, and O$_2$ at 0.76 \mm, and in the IR O$_3$ at 9.6 \mm, CH$_4$ at 7.6 \mm\ and N$_2$O at 17 \mm. These spectral features are shown in detail in Figure~\ref{multi_evol} (top: visible to NIR 0.4 to 3 \mm, bottom IR 3 to 20 \mm) for the planetary evolution models for a cooling WD host.

\section{Discussion \& Conclusion}

Due to the similarity in size to Earth and relatively stable environments, WDs are very interesting places to search for and characterize Earth-sized HZ planets. The contrast ratio between Earth-sized planets and their small WD hosts are several orders of magnitudes higher than for an Earth-sized planet around a main sequence star, however no such planet has been found yet. Finding and characterizing such planets in the HZ of WDs is challenging for several reasons. Due to the small size of WDs, such planets have a significantly lower transit probability than around main-sequence stars' HZs \citep{loeb13}. Although transits of HZ planets would occur very frequently ($<$10 hr orbital periods) their transit would only last a few minutes. It is also important to note that the close-in separation of the WD HZ would cause planets to experience extreme tidal forces. \cite{vera19b} finds that often Earth-sized planets orbiting at less than 10 times their WD's Roche limit would tidally disrupted, implying low chances of survival for the planetary orbits described in this paper ($a/r_{roche}$ = 2 - 4.5, Table~\ref{model_summary}).  In addition, cool white dwarfs have a very low flux due to their small size, limiting our search for such planets to nearby targets.

However finding rocky planets in the HZ of WDs would allow first insights into the composition and environment of such second-generation planets and the intriguing question whether they could support life. Here we present high-resolution transmission spectra for Earth-like planets orbiting WDs at the Earth-equivalent distance as well as explore how a rocky planet in the HZ of a WD  would evolve through the several billion year timescale when the WD host would provide a stable temperature environment. 

While finding rocky planets in the HZ of a WD is extremely challenging, we shows that upcoming telescopes could characterize the atmospheres of such potentially Earth-like planets, if they exist. All high-resolution transmission spectra will be available online upon publication of this study and can be used as a tool to prepare and interpret upcoming observations with JWST, the Extremely Large Telescopes, as well as mission concepts like Origins, HabEx, and LUVOIR.

\acknowledgements

This work was supported by the Carl Sagan Institute at Cornell University.

\end{document}